\pdfoutput=1

\documentclass[11pt]{article}
\usepackage{jheppub}

\usepackage{amsmath}
\usepackage{amssymb}
\usepackage{graphicx,color,slashed}
\usepackage{bbm} 
\usepackage{ifpdf}

\newcommand{\cO}{\mathcal{O}}

\newcommand{\be}{\begin{equation}}
\newcommand{\ee}{\end{equation}}
\newcommand{\ba}{\begin{eqnarray}}
\newcommand{\ea}{\end{eqnarray}}

\newcommand{\lp}{\left(}
\newcommand{\rp}{\right)}

\newcommand{\N}{\mathcal{N}}

\def\K{{K\"{a}hler} }

\title{de Sitter extrema and the swampland}

\author{Christoph Roupec,}
\author{Timm Wrase}

\affiliation{Institute for Theoretical Physics, TU Wien,\\
Wiedner Hauptstrasse 8-10/136, A-1040 Vienna, Austria}

\emailAdd{christoph.roupec@tuwien.ac.at}
\emailAdd{timm.wrase@tuwien.ac.at}

\notoc

\abstract{

\noindent
Recently it has been conjectured that string theory does not allow for dS vacua or dS extrema. To scrutinize such a conjecture, it is important to study concrete string theory compactifications and spell out their assumptions and potential shortcomings. We do so for one particular class of string compactifications, namely classical compactifications of type II string theory with fluxes, D-branes and O-planes on manifolds with SU(3) structure. In particular, we revisit previously found dS critical points, construct many new ones and check whether they invalidate the dS swampland conjecture.
}

\begin{document}

\maketitle

\newpage

\section{Introduction}
String theory provides us with a tractable UV complete theory of quantum gravity that has led to deep insights into what is possible and what is not possible in a theory of quantum gravity. This has manifested itself in two concepts, the gigantic string landscape of consistent low energy theories that arise from string compactifications and the even larger swampland of seemingly consistent low energy theories that, however, cannot be consistently coupled to gravity. In order to distinguish between the two, certain so called swampland conjectures have been formulated (see for example \cite{Brennan:2017rbf} for a recent review with many references). While some of these conjectures have been firmly established in string theory, like the absence of global symmetries, other conjectures are more speculative.

One of the heavily debated conjectures in the string community is related to the status of dS vacua in string theory. Based on the seminal KKLT \cite{Kachru:2003aw} and LVS \cite{Balasubramanian:2005zx} constructions, there has been a lot of activity in constructing dS vacua and theories of inflation in string theory, see for example \cite{Baumann:2014nda} for a recent review. However, the original KKLT and LVS scenarios have also been criticized throughout the years (see \cite{Greene:2015fva, Danielsson:2018ztv} for details and references). This criticism has cumulated into the recently stated dS swampland conjecture \cite{Obied:2018sgi} \footnote{This conjecture has been further studied and discussed in \cite{Agrawal:2018own, Agrawal:2018mkd, Dvali:2018fqu, Andriot:2018wzk, Banerjee:2018qey, Aalsma:2018pll, Achucarro:2018vey, Garg:2018reu, Lehners:2018vgi, Kehagias:2018uem, Dias:2018ngv, Denef:2018etk, Colgain:2018wgk, Paban:2018ole}.}
\be\label{eq:swamp}
|\nabla V| = \sqrt{g^{ij} \partial_{\phi^i} \partial_{\phi^j} V} \geq c \cdot V\,.
\ee
Here the scalar potential $V$ is a function of the scalar fields $\phi^i$ that arise in any given string theory compactification and the constant $c$ is conjectured to be $\mathcal{O}(1)$. Note that the above condition does not only exclude meta-stable dS vacua but also unstable dS critical points (i.e. saddle points) for which $|\nabla V|=0$ and $V\geq 0$. 

The authors of \cite{Obied:2018sgi} do not undertake the admittedly fairly difficult task of going through every dS construction in the string theory literature and show how or why it could be compatible with their dS swampland conjecture. They rather provide a set of explicit examples that almost all seem to satisfy a bound of the above form. In particular, the largest class of examples in \cite{Obied:2018sgi} that support the above conjecture are classical flux compactifications of type II string theory. These setups are believed to give rise to scalar potentials with dS critical points, first found in \cite{Caviezel:2008tf, Flauger:2008ad}, which seems to be in tension with the above conjecture.

In this paper we reexamine and construct new dS critical points in classical flux compactifications of type II supergravity. We spell out all the assumptions that go into their construction and check whether they can be made consistent with string theoretical constraints imposed onto the type II supergravity, like flux quantization, large volume and small string coupling.

The outline of the paper is as follows: In section \ref{sec:setup} we discuss the setup of interest and discuss existing no-go theorems similar to the dS swampland conjecture. Then we spell out the ugly details of these constructions in section \ref{sec:shortcomings} and discuss whether they are compatible with embedding these solutions in full fledged string theories. In section \ref{sec:dScrit} we discuss the existing dS critical points that were found so far, as well as new ones, and we study whether they are compatible with flux quantization, realistic tadpole conditions and the requirement of large volume and weak coupling. We summarize our results in section \ref{sec:conclusion}.

\section{Classical type II flux compactifications}\label{sec:setup}
In this section we review the details and ingredients used in flux compactifications of type II string theory or rather its low-energy limit type II supergravity. We spell out some of the potential ingredients that can be used and discuss the status of obtaining vacua with different values of the cosmological constant. Here a particular role is played by some simple no-go theorems that are satisfied by these setups in certain cases.

In this paper we are restricting to compactifications of 10d type II supergravity on manifolds with SU(3) structures that give rise to $\N=1$ supergravity theories in four dimensions. In order to generate a classical scalar potential we can turn on RR- and NSNS-fluxes and include for example orientifolds, D-branes, NS5-branes and KK-monopoles.

A particular setup that is very interesting is massive type IIA compactified on Calabi-Yau manifolds. The scalar potential arises from RR-fluxes $F_0$, $F_2$, $F_4$,\footnote{In order to give rise to a maximally symmetric spacetime, only the $F_4$-flux can fill the external space. This is often described in terms of the dual $F_6 = \star F_4$ threading the internal manifold.} NSNS-flux $H_3$ and O6-planes. The generic reduction from ten to four dimensions was worked out in \cite{Grimm:2004ua}. The scalar potential was minimized and the resulting AdS vacua were studied in \cite{Villadoro:2005cu, DeWolfe:2005uu, Camara:2005dc, Ihl:2006pp}. The scalar potential in these setups depends on all moduli except the $C_3$-axions, furthermore, the $F_4$-flux is unconstrained by tadpole conditions. In the limit of large $F_4$-flux one obtains a parametrically separation between the AdS and the KK scale as well as parametrically large volume and weak coupling, so these AdS vacua are very well controlled.

The existence of these well controlled AdS vacua led to the numerical search \cite{Hertzberg:2007ke} for flat regions with positive values of the scalar potentials in the three explicit models \cite{Villadoro:2005cu, DeWolfe:2005uu, Ihl:2006pp}. This search did not find regions with $V>0$ and small slow-roll parameter
\be
\epsilon=\frac{1}{2}\frac{|\nabla V|^2}{V^2}\,.
\ee 
Note that this $\epsilon$ parameter is related to the constant $c$ in the dS swampland conjecture in eqn. \eqref{eq:swamp} via $\epsilon=c^2/2$.

One can actually show analytically \cite{Hertzberg:2007wc} that regions with small $\epsilon$ cannot exist in these scalar potentials. If we define the volume modulus $\rho$ and the dilaton modulus $\tau$ via
\be
\rho = (vol_6)^\frac13\,, \qquad \quad \tau = e^{-\phi} \sqrt{vol_6}\,,
\ee
then the scalar potential has the following form
\be
V(\rho,\tau) = \frac{A_H}{\rho^3 \tau^2} + \sum_{p=0,2,4,6} \frac{A_p}{\rho^{p-3} \tau^4} - \frac{A_{O6}}{\tau^3}\,,
\ee
where the $A$'s are positive functions of the other moduli and the flux quanta. One now finds that 
\be
-\rho \frac{\partial V}{\partial \rho} - 3 \tau \frac{\partial V}{\partial \tau} = 9V+\sum_{p=0,2,4,6} p \frac{A_p}{\rho^{p-3} \tau^4}\geq 9V\,.
\ee
From the above inequality it is clear that dS critical points are forbidden in these models since for a dS critical point the left-hand-side would have to vanish, while the right-hand-side would be positive. It is also straightforward to derive an explicit bound on $\epsilon$ and therefore $c$ in these models and one finds $\epsilon =c^2/2 \geq 27/13$ \cite{Hertzberg:2007wc}, which implies that also slow-roll inflation is not possible in this class of models.

The existence of this no-go theorem motivated people to look for broader classes of string compactifications \cite{Saltman:2004jh, Silverstein:2007ac, Haque:2008jz, Flauger:2008ad, Caviezel:2008tf, Danielsson:2009ff, deCarlos:2009fq, deCarlos:2009qm, Caviezel:2009tu, Danielsson:2010bc, Dong:2010pm, Andriot:2010ju}.\footnote{Within this particular setup the scalar potential is generated at tree-level and involves essentially all moduli, so it probably does not make sense to invoke quantum corrections. However, if one turns off certain fluxes so that some moduli do not appear in the tree-level scalar potential, then one can invoke quantum corrections and mirror for example the LVS construction in type IIA \cite{Palti:2008mg}.} One generalization that was extensively studied and also prominently featured in the examples provided in support of the dS swampland conjecture is the inclusion of curvature. We can go beyond Ricci flat $CY_3$ manifolds and for example study more general manifolds with SU(3) structure and non-vanishing scalar curvature. Additional  ingredients that have appeared are fractional Chern-Simons forms, NS5-branes and KK-monopoles \cite{Silverstein:2007ac}. 

A priori one is also not restricted to compactifications of type IIA any more. One can now study type IIB as well since curvature together with other ingredients can generate interesting tree-level scalar potentials that can depend on all moduli. This opens up the possibility of studying a whole zoo of models and of excluding some classes of them based on no-go theorems similar to the one above \cite{Silverstein:2007ac, Caviezel:2008ik, Haque:2008jz, Flauger:2008ad,  Danielsson:2009ff, Caviezel:2009tu, Wrase:2010ew, Andriot:2016xvq, Andriot:2017jhf}. All of these no-go theorems provide bounds with $c\gtrsim \mathcal{O}(1)$ and are therefore consistent with the dS swampland criterion. However, they were originally invented in order to identify a minimal set of ingredients that evades any such no-go theorem. Probably the simplest and most studied class of such models is the compactification of (massive) type IIA with fluxes, O6-planes and curvature.\footnote{The inclusion of for example D6-branes is trivial and does not improve the situation since the extra term $V_{D6} = +\frac{A_{D6}}{\tau^3}$ would simply combine with the O6-plane term. If $A_{D6} > A_{O6}$, then there is trivially no solution since all contributions to the scalar potential are positive and scale with negative powers of the dilaton and volume modulus.} In this case the scalar potential is of the above form with one extra term arising from the non-trivial curvature of the internal manifold
\be
V(\rho,\tau) = \frac{A_H}{\rho^3 \tau^2} + \sum_{p=0,2,4,6} \frac{A_p}{\rho^{p-3} \tau^4} - \frac{A_{O6}}{\tau^3}+\frac{A_{R_6}}{\rho \tau^2}\,.
\ee
For spaces with negative curvature we have $A_{R_6}>0$ and one can show that no no-go theorem similar to the one above exists. Here it is important to stress (as was done in \cite{Obied:2018sgi}) that \emph{no no-go} in the $(\rho,\tau)$-plane does not imply \emph{go}. It is still possible that other moduli prevent the existence of dS critical points. In order to show that there can \emph{not} be such a simple scaling argument of the scalar potential that excludes dS critical points one has to analyze the scalar potential for explicit models and check whether they allow for dS solutions. This has been done and the first dS critical points in this class of models were constructed in \cite{Caviezel:2008tf, Flauger:2008ad}. These constructions prove that there cannot be a simple no-go theorem like the one above that forbids dS critical points for all values of $\rho$ and $\tau$ as well as for all values of the $A$'s. However, this does not disprove the dS swampland conjecture. The reason for this is that a trustworthy string theory solution, derived in the supergravity limit, has to satisfy many additional constraints like for example tadpole cancellations, flux quantization and the absence of large $\alpha'$ and string loop corrections. We will discuss all such potential issues in general in the next section before looking at concrete dS critical points in section \ref{sec:dScrit}.

\section{Potential shortcomings of the setups}\label{sec:shortcomings}
In this section we expose the potentially ugly details and discuss shortcomings that might or might not be problematic, when one wants to identify the solutions of the 4d scalar potential with solutions of a full string theory. We hope that this section is helpful in identifying what features trustworthy string compactifications should have and how one could improve the existing constructions of dS critical points in these classical setups.

\subsection{What are the relevant scalars?}\label{ssec:moduli}
For generic SU(3) structure manifolds that are not Ricci flat and therefore no Calabi-Yau manifolds, it is not clear anymore what the lightest fields are \cite{KashaniPoor:2006si, KashaniPoor:2007tr}. Therefore, we cannot easily write down a low energy effective action and study the corresponding scalar potential. However, it was shown in \cite{Cassani:2009ck} that for compactifications on group manifolds one can restrict to expanding in forms that are left-invariant with respect to the action of the group. This gives rise to a consistent truncation so that solutions obtained in four dimensions lift to solutions of the ten dimensional equations of motion. We include in our setups also an orientifold projection but this should not change the above statements (except for one caveat that we discuss in detail in the next subsection). 

While it has not been strictly proven that after the orientifold projection we still have a consistent truncation, it is probably a straightforward exercise to show this. Furthermore, it was explicitly shown that some of the dS critical points that were found in four dimensions lift to ten dimensions \cite{Danielsson:2010bc}. So it seems that the expansion in left-invariant forms is a very useful and valid approach. However, note that generically the `moduli' in the resulting four dimensional theories are not necessarily the lightest fields and they could potentially be even heavier than the KK scale, which was pointed out in \cite{Andriot:2018wzk}. However, this is not a problem since we satisfy the ten dimensional equations of motion and thus have found a valid solution. These 10d solutions have so far one shortcoming that we discuss in the next subsection.

\subsection{Integrated equations of motion for intersecting sources}\label{ssec:smeared}
Generically all of these setups of interest include intersecting sources in which case it seems not possible to directly solve the ten dimensional equations of motion.\footnote{In the case of parallel stacks of sources one can solve the equations of motion by introducing an appropriate warp factor. The same is true for branes within branes but as soon as two sources intersect there is (as far as we know) no known solution.} Since one cannot exactly solve the equations of motion for such sources, people have restricted themselves to solving only the integrated equations of motion that do not have delta functions in them. This procedure is often called `smearing' of the sources, which has sometimes caused confusion because localized objects like a single O6-plane can of course not be smeared. 

Solving the integrated equations of motion is a necessary but potentially not sufficient requirement for a localized solution to exist. In \cite{Douglas:2010rt} it was pointed out that there exists the possibility of the absence of localized solutions despite the existence of the smeared solution. This was further studied in \cite{Blaback:2010sj, Blaback:2011nz, Blaback:2011pn, Saracco:2012wc, McOrist:2012yc} but it is still unclear whether dS solutions that were found by solving the integrated equations of motion have corresponding fully localized solutions. Note that all dS critical points that were found in type IIA require the mass parameter to be non-zero, so they cannot be lifted to M-theory. All existing dS critical points also have intersecting sources but refined no-go theorems for parallel sources do not forbid dS vacua with only parallel sources \cite{Andriot:2016xvq}. If one could find a dS solution with only parallel sources, then one can solve the localized equations of motion by introducing warp factors. So this seems to be an interesting avenue to pursue in order to make progress on this important issue.

\subsection{Blow-up modes from orbifold singularities}
All of the existing dS critical points in these classical setups are obtained for compactifications on orbifolds of group manifolds (or so called twisted tori). For the standard Abelian orbifolds it was shown in \cite{Flauger:2008ad} that refined no-go theorems exist that forbid dS critical points for all but the $\mathbb{Z}_2 \times \mathbb{Z}_2$ orbifold. In \cite{Danielsson:2011au} some non-Abelian orbifolds were studied and dS critical points were constructed for $\Delta(12)$, which contains $\mathbb{Z}_2 \times \mathbb{Z}_2$ as a subgroup. 

All of these orbifold constructions are very similar to the probably more familiar cases of flux compactifications on toroidal orbifolds. In that case, and in the more general setup with curvature, one has to deal with the orbifold singularities. This has only been addressed explicitly in the toroidal orbifold examples \cite{DeWolfe:2005uu, Ihl:2006pp}, which only give rise to AdS vacua. There the authors showed how one can remove the orbifold singularities via blow-ups. This introduces new cycles that give rise to additional moduli and can also support additional fluxes. It was possible to stabilize all additional moduli from the twisted sector at a smaller scale. Therefore, they did not affect the properties of the leading bulk AdS solutions in these examples. Due to such a scale separation between the bulk and blow-up moduli, one is tempted to suspect that the same holds for orbifolds of group manifolds that give rise to dS critical points. However, so far this has not been checked for any concrete example. Another subtlety is that the bulk moduli are left invariant forms under the group action (see \ref{ssec:moduli} above). For the blow-up modes this is not true so we are not dealing with a consistent truncation anymore.

\subsection{The mass parameter in type IIA}
In all these flux compactifications of type IIA the Romans mass parameter $m_0=F_0$ \cite{Romans:1985tz} plays an important role. Due to the presence of this mass parameter one might be worried that we cannot understand these supergravity solutions within perturbative string theory and that it is unclear how to define the O6-planes. The mass parameter and the corresponding absence of an M-theory lift seem problematic and might prevent us from better understanding even the supersymmetric AdS solutions \cite{Banks:2006hg}. However, it was also argued that massive type IIA string theory can never be strongly coupled in weakly curved regions of spacetime \cite{Aharony:2010af}, which is a nice feature that deserves further study in the context of these flux compactifications. 

While there is certainly a lot of interesting physics associated with the mass parameter in type IIA, we simply point out that it is not crucial for the existence of classical dS critical points. It was shown in \cite{Caviezel:2009tu} that one can formally T-dualize the massive type IIA setup with O6-planes and curvature. The dual theories are type IIB supergravity compactified on SU(2)-structure manifolds with O5- and O7-planes.\footnote{Strictly speaking the T-dual setup can have so called non-geometric fluxes which we turn off. We have however the freedom of turning on some new geometric fluxes that would be non-geometric in the dual type IIA setup.} These SU(2)-structure manifolds, which in the simplest cases, are closely related to the SU(2)-holonomy space $T^2 \times K3$, have in particular 1-cycles and 5-cycles, so we can turn on the RR-flux $F_1$. This flux is the T-dual of the Romans mass parameter $F_0$. These compactifications can give rise to supersymmetric AdS vacua with parametrically large volume and weak string coupling \cite{Caviezel:2009tu}. In that paper the authors also found one model that gave rise to a dS critical point. All type IIB setups studied so far have intersecting O5- and O7-planes and one can again only solve the integrated equations of motion. It would be interesting to check whether one can understand such setups in F-theory in order to make progress on the issue discussed in subsection \ref{ssec:smeared}.

Summarizing, the type IIA mass parameter is not crucial in order to obtain dS critical points, since they have also been found in T-dual type IIB setups.

\subsection{Flux quantization}
Flux quantization is a requirement that the underlying string theory imposes on the low energy supergravity. However, in the search for AdS, Minkowski or dS vacua it is not always imposed from the beginning since it is much easier to numerically find solutions, if one lets the fluxes take continuous values. Often it is the case that solutions exist for large ranges of fluxes that contain many correctly quantized fluxes so that taking flux quantization into account does not really affect the existence of solutions. However, in the search for dS solutions it has been noticed that they often seem to appear only for a very restricted set of fluxes and the possibility exists that this restricted set does not contain properly quantized fluxes. So it is worthwhile to reexamine this issue.

The NSNS $H_3$-flux quantization condition is simply the requirement that the integral of the $H_3$-flux over all cycles in integer homology has to give an integer. The RR-flux quantization is more subtle in the presence of non-vanishing $H_3$-flux and one should use in principle $H_3$-twisted K-theory \cite{Ktheory1, Ktheory2}. However, it was shown in \cite{twistedhom} that for simply-connected six manifolds the $H_3$-twisted K-theory is isomorphic to the $H_3$-twisted cohomology. So in order to properly quantize the RR fluxes one can simply calculate the $H_3$-twisted (co-) homology. Since the search for dS vacua in classical type II flux compactifications has only led to phenomenological uninteresting dS critical points, this exercise has only been carried out for the compactification on the group space $SU(2)\times SU(2)=S^3 \times S^3$ in \cite{Danielsson:2011au}. Given the recent dS swampland conjecture it might be worthwhile to do this calculation for other manifolds as well.

Another interesting approach to obtain correctly quantized NSNS $H_3$-flux as well as many different interesting compact geometries is called the base-fiber splitting. Here one starts with $T^6$ and splits  it into a base $T^n$ and a fiber $T^{(6-n)}$. Then one chooses for each circle in the base a matrix $M_i \in \mathfrak{so}(6-n,6-n)$, $i=1,\ldots, n$. The matrix entries correspond to $H_3$-fluxes, geometric fluxes and even non-geometric so called $Q$-fluxes, which we can take to be zero. The flux quantization conditions amount to the requirement that $e^{\lambda^i M_i} \in SO(6-n,6-n;\mathbb{Z})$ for all $\lambda^i \in \Lambda$, where $\Lambda$ is the torus lattice $T^{n}=\mathbb{R}^n/\Lambda$. See \cite{Bergman:2007qq} for some more mathematical aspects of this approach as well as a list of references and \cite{Cvetic:2007ju, Ihl:2007ah, Flauger:2008ad, Caviezel:2009tu} for some worked out examples in the context of type II flux compactifications.

In the case of flux quantization for SU(3)-structure manifolds we have to keep in mind the discussion in subsection \ref{ssec:moduli}. We are expanding our fields and fluxes in terms of left-invariant forms that are not necessarily in twisted cohomology . If they are not in twisted cohomology then it appears that they do not have to be quantized. However, if the fluxes appear in the tadpole conditions then they have to cancel the quantized charge of localized sources, which still imposes constraints.

To summarize, flux quantization has not been properly implemented in all examples that give rise to classical dS critical point. It can be worked out for orbifolds of group or coset spaces using the method outlined in appendix D of \cite{Danielsson:2011au}.

\section{dS critical points}\label{sec:dScrit}
In this section we analyze dS critical points that were found numerically in the context of type IIA flux compactifications on SU(3)-structure manifolds and we discuss many new dS critical points that we found. We are particularly interested in their compatibility/incompatibility with the dS swampland conjecture.

So far people studied type IIA orbifold compactifications on six real dimensional group spaces $G$. For an orbifold group that is a subgroup of SU(3) one then obtains an $\N=2$ theory in four dimensions. Doing a further orientifold projection that gives rise to O6-planes, one obtains a four dimensional supergravity theory with $\N=1$ supersymmetry. This case is fairly similar to the case of orbifold compactifications on $T^6$, except that the more general group spaces have non-vanishing curvature. The scalar potential in four dimensions arises generically from F- and D-terms \cite{Grimm:2004ua, Robbins:2007yv}. All the details of these constructions have been reviewed in \cite{Danielsson:2011au}, to which we refer the interested reader.

The resulting four dimensional scalar potentials in these setups depend on up to 7 complex scalar fields. These include in particular the dilaton and geometric scalars arising from expanding the holomorphic 3-form $\Omega$ and the \K form $J$. The parameters of the scalar potential arise from the $H_3$-fluxes, the $F_p$-fluxes for $p=0,2,4$ and some so called geometric fluxes that encode the curvature of the SU(3)-structure manifold. The dS critical points are found by numerically minimizing the rather complicated scalar potential not just with respect to the moduli but also with respect to the flux parameters. The first three such dS critical points were discovered in \cite{Caviezel:2008tf, Flauger:2008ad}. A more systematic study of different group manifolds has led to more than a dozen different compactification spaces that can give rise to dS critical points \cite{Danielsson:2011au}. For these more than a dozen examples, slightly more than one hundred different critical points have been found in \cite{Danielsson:2012et}. We have been able to find more than 300 new dS critical points in these examples so that we have a large set of over 400 dS critical points that we analyze in this section.

Before we discuss these critical points in detail, let us point out what seems like obvious shortcomings: Since we allow the fluxes to vary continuously these dS critical points do not satisfy the flux quantization conditions. Some fluxes also appear in the tadpole condition $dF_2 + F_0 H_3 = j_{O6/D6}$ so they also do not satisfy the tadpole condition. While this seems fatal, this is not really the case. Each given model has a lot of symmetries and we can rescale the moduli and fluxes while still staying at the critical point. So for each model we can work out the correct flux quantization condition and its scaling symmetries and then use these to satisfy the flux quantization and/or the tadpole condition. If the scaling symmetries are not sufficient for this, then we can also start searching for dS critical points in the neighborhood of the existing critical points for this model. Usually there is a region in parameter (i.e. flux) space for which dS critical points exist. So these dS critical points provide a good starting point for finding correctly quantized examples that satisfy the proper tadpole condition. However, working out the correct quantization condition for each of the more than a dozen examples is tedious and mapping out the complete higher dimensional parameter space is extremely time consuming, so we leave these tasks for the future.

For the most symmetric and simplest compactification space that gives rise to dS critical points, namely the compactification on $S^3 \times S^3/\mathbb{Z}_2 \times \mathbb{Z}_2$, the parameter space has been mapped out. In this example, all flux parameters except one can be set to unity by rescaling the moduli, so that the scalar potential effectively only depends on one single parameter. For a certain range of this parameter, it was found that the scalar potential has dS critical points \cite{Danielsson:2010bc}. Leaving that range to one side simply leads to Minkowski and AdS vacua while the other end of the range corresponds to a degenerate scalar potential that develops a pole. For this particular example, the correct flux quantization conditions were worked out in \cite{Danielsson:2011au}. There it was found that imposing the correct quantization conditions for the mass parameter and the $H_3$-flux is sufficient to show that all dS critical points are at small volume and at strong coupling. Thus, all these dS critical points receive large $\alpha'$ and string loop corrections and should not be trusted. 

Given the above result we initiate here the study of all the dS critical points in the roughly one dozen other examples. First of all we notice that the simple numerical minimization that we implemented in Mathematica gives generically solutions with values for the fluxes and moduli that are spread around one. While there is a substantial variation, the volume of the internal space is not necessarily small, being  of $\cO(10)-\cO(100)$ in natural units, and the inverse string coupling seems usually acceptable with $e^{-\phi} \sim \cO(10)$. However, these examples generically also have tadpole contributions from the fluxes that are positive and negative with absolute value of $\cO(10)$. While we can add in principle D6-branes to get arbitrarily large positive contributions to the tadpole, we cannot change the number of O6-planes which is usually of order one. These O6-planes and their negative contributions to the tadpole conditions is what circumvents the no-go theorem by Maldacena-N\'u\~{n}ez \cite{Maldacena:2000mw}. Thus, all dS critical points necessarily have one or more tadpoles with negative contributions and these negative contributions should be order one not order ten.

In order to fix these too large negative tadpole conditions we have identified a simple universal scaling symmetry that exists for all models. In particular we have rescaled the \K moduli by a constant and the complex structure moduli by the inverse constant. Demanding that the scalar potential only changes by an overall factor under this rescaling fixes the scaling of all the fluxes. The metric fluxes which are model dependent stay invariant under this rescaling but the tadpole and the mass parameter transform non-trivially. In these setups there are four different tadpole conditions corresponding to four different left-invariant 3-forms. We have used the above scaling symmetry to change all dS critical points so that the most negative flux contribution in these four tadpole conditions corresponds to a single O6-plane. The other tadpole conditions have then larger flux contributions that could potentially be accounted for by a single O6-plane plus additional D6-branes. After this rescaling the number of models with large volume and weak string coupling has substantially decreased but there are still many models that satisfy this first necessary requirement. However, for all these models we noticed that a large volume correlates with a small mass parameter $F_0$. In particular, $F_0^2 \cdot vol_6$ takes values between $\cO(1)$ and $\cO(100)$, so that a very large volume implies an $F_0$ flux that is smaller than unity.\footnote{Given this numerical observation it would be very interesting to check whether such a correlation can be proven analytically. } The quantization condition for $F_0$ is fairly simple and it has to be integer quantized, which is in tension with the requirement of a large volume. 

Given the above results, it seems clear that these models deserve further attention. We have found that models with realistic tadpole conditions, $vol_6 \gtrsim \cO(100)$ and $e^{-\phi} \gtrsim \cO(10)$ exist. These might be in tension with the dS swampland conjecture. However, in order to say this for sure, one has to impose further conditions like the correct quantization of all fluxes. It might also be desirable to obtain an even larger volume and smaller string coupling to be certain that $\alpha'$ and string loop corrections do not invalidate the solution.

We like to quickly compare our findings above with the properties of AdS vacua in these setups. As we mentioned above for Calabi-Yau flux compactifications (i.e. in the absence of curvature) the large $F_4$ flux limit gives rise to AdS vacua with arbitrarily large volume and small string coupling \cite{DeWolfe:2005uu}. Similar statements are not known for compactifications on generic SU(3)-structure manifolds. However, concrete examples with AdS solutions that exhibit such a scaling limit do exist. For dS vacua such a scaling limit has not been found and our results above are consistent with the absence of such a limit. However, the above results might also be due to the absence of an analytic understanding of dS solutions. Therefore, we performed a numerical search in these models for AdS vacua. We found for the AdS solutions that, after rescaling the tadpole conditions to realistic values, $F_0^2 \cdot vol_6$ takes similarly values between $\cO(1)$ and $\cO(100)$. So the numerical analysis of AdS vacua did likewise not give rise to very large volumes with properly quantized $F_0$ flux, although such solutions are known to exist from analytic studies.  

In the T-dual context of type IIB flux compactifications only one dS critical point was presented in \cite{Caviezel:2009tu}. This dS critical point had two tachyonic directions, one of which had large $\eta$ value of $\eta \approx -3$. This critical point was not at large volume and small string coupling and the fluxes were not properly quantized. However, this proved that dS critical points do not only exist in massive type IIA. This type IIB setup has not really been studied much in the literature and certainly deserves further study.

\section{Conclusion}\label{sec:conclusion}
We have discussed the fairly simple and seemingly well controlled class of classical type II flux compactifications. In particular, we revisited the existing results in the literature that are potentially in conflict with the recently proposed dS swampland conjecture \cite{Obied:2018sgi}. As a first step towards understanding such a potential tension, we have spelled out the problematic details of these setups. Here two points are particularly important: 1) All dS critical points that have been found up to date involve intersecting O-planes and therefore one cannot solve the ten dimensionally equations of motion pointwise but only after integration over the internal space. 2) The existing critical points were all found numerically and they have not been shown to exist in a large volume and weak coupling regime, for properly quantized fluxes.\footnote{For one particular example a 1-parameter family of dS solutions was studied in great detail and it was found that imposing the proper flux quantization forces all dS critical points to be at small volume and large string coupling, so that they are not trustworthy \cite{Danielsson:2011au}.}

Here we have taken first steps towards checking point 2). We have taken all existing over one hundred dS critical points and generated more than 300 new ones. These critical points arise in massive type IIA compactifications on more than one dozen different compactification spaces. By construction they do not satisfy flux quantization and the correct integrated tadpole $\int dF_2 + F_0 H_3 = N_{D6}-2N_{O6}$ for integers $N_{D6}$ and $N_{O6}$. However, all the models have symmetries that allow us to rescale the moduli and fluxes. Unfortunately, these symmetries and the flux quantization conditions are model dependent and need to be worked out carefully for each model, which we leave to the future. We restricted ourselves to identifying one universal rescaling that allowed us to rescale the tadpole such that it satisfies $\int_{\Sigma_3} dF_2+F_0 H_3 \geq -2$ for all 3-cycles $\Sigma_3$. We found that after this rescaling all dS critical points had $(F_0 )^2 vol_6$ of at most $\cO(100)$. Since the mass parameter $F_0$ has to be integer quantized, there is a potential tension with the requirement of being at large volume in order to suppress $\alpha'$ corrections. The inverse string coupling could be of $\cO(10)$, which might be sufficient to suppress string loop corrections. Based on these results we believe that these dS critical points deserve further study because they might provide the simplest explicit string theory setups that carry the potential of falsifying the dS swampland conjecture.

Given the existence of these dS critical points and a tension between the conjecture and the Higgs potential \cite{Denef:2018etk}, one might also entertain a modified dS swampland conjecture \cite{Andriot:2018wzk, Garg:2018reu, Denef:2018etk} that does not only constrain the $\epsilon$ slow-roll parameters (cf. eqn. \ref{eq:swamp})
\be
\epsilon = \frac12 \lp\frac{V'}{V} \rp^2 \geq \frac{c^2}{2} \sim \cO(1)\,,
\ee
but also the $\eta$ slow roll parameter. Here we like to recall, that the $\eta$ parameter is usually larger than unity for all the classical dS critical points discussed in this paper (see however \cite{Blaback:2013fca} for slightly smaller values). Similar to the no-go theorems we discussed in section \ref{sec:setup} that prevent vanishing $\epsilon$ for some ingredients, there exist no-go theorems against stability for certain setups \cite{Shiu:2011zt, Danielsson:2012et, Junghans:2016uvg, Junghans:2016abx}.

Fully stable classical dS vacua can be obtained when using so called non-geometric fluxes \cite{deCarlos:2009fq, deCarlos:2009qm, Danielsson:2012by, Blaback:2013ht, Damian:2013dq, Blaback:2015zra}. These non-geometric fluxes appear naturally in string theory, for example, if one T-daulizes two directions of a $T^3$ threaded with $H_3$-flux. However, at the level of supergravity it is unclear whether they allow for a limit in which $\alpha'$ corrections are absent. Naively this should never be the case since non-geometric fluxes seem to require some cycles to be of string length. However, one can T-dualize geometric solutions at arbitrarily large volume and weak coupling to find solutions with non-geometric fluxes that should not receive significant corrections since their T-dual geometric counterpart does not receive corrections either. It would be interesting to understand this better and determine when one can trust supergravity solutions that involve non-geometric fluxes.

In addition to the relatively simple setups discussed in this paper, there are many more approaches to the construction of classical dS vacua (see for example \cite{Silverstein:2001xn, Maloney:2002rr, Saltman:2004jh, Silverstein:2007ac, Haque:2008jz, Danielsson:2009ff, deCarlos:2009fq, deCarlos:2009qm, Dibitetto:2010rg, Dong:2010pm, Andriot:2010ju, VanRiet:2011yc, Danielsson:2012by, Blaback:2013ht, Damian:2013dq, Dodelson:2013iba, Blaback:2015zra}). These should also be carefully analyzed in order to see whether they could be compatible with the dS swampland conjecture. These constructions include extra ingredients, which makes them \emph{more} generic, or they arise in other dimensions or they involve the non-geometric fluxes discussed above. Another, potentially more complicated goal, would be to reach an agreement within our community on the consistency or potential shortcomings of the existing \emph{quantum} dS vacua scenarios like KKLT \cite{Kachru:2003aw} and LVS \cite{Balasubramanian:2005zx}.

\acknowledgments
We are grateful to D.~Andriot, U.~Danielsson, G.~Shiu, T.~Van~Riet for enlightening discussions and comments on an earlier version of this paper.  We thank L.~L\"uftinger for running some of the calculations on his computer. This work is supported by an FWF grant with the number P 30265.

\bibliographystyle{JHEP}
\bibliography{refs}

\end{document}